\documentstyle[preprint,aps,eqsecnum]{revtex}
\begin{document}
\title {Quantum transport of composite fermions in narrow constrictions}

\author {D.V. Khveshchenko}

\address {NORDITA, Blegdamsvej 17, Copenhagen DK-2100, Denmark}

\maketitle

\begin{abstract}
\noindent
We elaborate on a single-mode description of the ballistic charge transport 
through a narrow constriction
in a compressible (composite fermion) Quantum Hall liquid. In the presence of 
long-range 
Coulomb interactions we find logarithmic deviations from the Luttinger liquid
behavior. These extra logarithmic factors are shown to
modify the conductance, shot noise spectrum, and other observables.

\end{abstract}
\pagebreak

It was first proposed by Wen \cite{1} to consider 
edge states of Fractional Quantum Hall Effect (FQHE) as a physical realization 
of the concept
of a 1D Luttinger liquid (LL).
Such hallmarks of the LL behavior as a power-law density of states
were predicted to be seen in direct tunneling experiments.
When, recently, such a long-awaited experiment became available \cite{2}, its 
results 
appeared to be rather unexpected and calling for a further 
refinement of the original theoretical description.

Namely, the authors of Ref.\cite{2} reported an approximate power-law behavior 
of the measured low-bias current-voltage characteristic
$I\sim V^{1/\nu}$ for tunneling from a metallic lead into a Quantum Hall (QH) 
state
at a continuously varying filling factor $\nu$. Taken at its face value, this 
experimental finding 
seemed to suggest that the most naive single-mode Littinger liquid picture 
might be applicable in a whole range of the probed values of $\nu$.

At first sight, such an implication would be difficult to reconcile with the
current phenomenology of the edge states.
Indeed, contrary to the experimental data of Ref.\cite{2} which show a 
remarkably 
smooth dependence on $\nu$, the available theoretical description is 
uncomfortably dependent on the precise (rational) value of $\nu$. In particular, 
 the very number of edge modes was predicted to vary drastically
from one value of $\nu$ to another, while the single-mode chiral Luttinger 
liquid description was only 
expected to be valid for atomically sharp edges
of the principal (Laughlin's) FQHE states at $\nu=1/(2n+1)$.

So far this theoretical picture has only been extended 
onto  Jain's FQHE states 
at $\nu=m/(2nm+1)$ whose edges are believed to support $m$ distinct modes.  
The exponent of the tunneling density of states characterizing  these fractions
was found  to be equal to $1+|2n+{1/m}|-|{1/m}|$ \cite{3}. 

Apart from that, the phenomenological theory offers no predictions for
neither incompressible fractions other than Jain's, nor any compressible ones, 
nor even for the case
of a constriction which is too narrow to physically accommodate multiple edge 
modes.

An appealing alternate approach that does not discriminate between 
incompressible and compressible
states was recently developed \cite{4} and shown to corroborate the results of 
Ref.\cite{3}. 
Thus its predictions also appear at odds with the experimental data from 
\cite{2}
for all $\nu\neq 1/(2n+1)$.

Since the findings of Ref.\cite{2}
were first reported, there have been several attempts of their rationalization 
\cite{5,6}. 
One way or another, these works addressed such a    
shortcoming of the phenomenological theory as a postulated, rather than derived,
form of coupling between the tunneling electron and the bosonic edge modes.
 
In the phenomenological approach, the form of this coupling is prescribed by the 
chiral 1D bosonization procedure
which represents individual FQHE edge channels in terms of chiral 1D bosons. By 
default, these channels 
are assumed to be all physically accessible for a tunneling electron, and, 
therefore, the 
corresponding chiral 1D bosons all contribute comparably to the bosonized 
expression for 
the electron operator.

On the contrary, the analyses of Refs.\cite{5} provided a host of evidence 
indicating 
 that the suppression of the tunneling density of states
is dominated by the only 
charge-carrying combination of the edge modes, that is the edge magnetoplasmon.

This observation is suggestive of a possibility to construct an effective 
single-mode
theory of the charge transport through a quantum point contact (QPC), 
regardless of the complexity of the detailed edge 
structure or even the very existence of well-defined edge modes \cite{6}.

In the present paper, we make another attempt in this direction and address the 
problem of ballistic transport through a QPC 
between (in general, different) compressible QH liquids at even denominator 
filling fractions. 

These states provide an interesting example of a 2D non-Fermi liquid "in 
disguise", 
an ultimate theoretical description of which still remains under construction.
However, according to the most recent
developments \cite{7}, at least at primary fractions $\nu=1/2p$ 
the nearly-Fermi-liquid quasiparticles named "composite fermions" (CFs)
form an ostensible Fermi surface at Fermi momentum $k_F=(4\pi n_e)^{1/2}$
and carry no electrical charge but the dipole moments only.
These quasiparticles experience strong residual interactions
leading, alongside with other effects,
to the diffusion-like pole in the density-density correlation function at small 
momenta $(q<<k_F)$ \cite{8}: 
$$\chi(\omega, {\bf q})=  {\sigma_qq^2\over i\omega + U_q\sigma_qq^2}  
\eqno(1)$$ 
where $\sigma_q$ is the momentum-dependent bulk conductivity
and $U_q=2\pi e^2/\epsilon_0q$ is the pairwise Coulomb potential with the 
dielectric constant 
$\epsilon_0$. Another distinct non-Fermi-liquid feature, for which both 
analytical 
\cite{9} and numerical \cite{10}
evidencies have been obtained, is a strong
enhancement of the density response at $2k_F$.

In what follows, we consider the regime of strong transmission (weak 
backscattering), complimentary
to that of weak tunneling which was experimentally probed in Ref.\cite{2}
and analyzed in Refs.\cite{5,6}. 

Provided the width of the constriction, which opens adiabatically into the   
2D regions, is comparable to the magnetic length $(\sim 10nm)$, there will be no 
transitions
between different transverse channels, and the electron motion inside the QPC 
will have an approximate 
1D character unaffected by the external magnetic field.

This allows one to introduce a single pair of transitive left-  and right-  
moving spinless fermion modes which adiabatically transform into dipole-like 
(neutral) bulk quasiparticles   
on either one or another side of the constriction. The explicit expression
for $\psi_{L,R}(x)$ in terms of the bulk fermion operators $\Psi_{L,R}({\bf r})$ 
will depend on the concrete geometry of the constriction.
For our purposes their kinetic energy can be taken in the usual form  
$$H_0= iv_F\int dx (\psi^{\dagger}_L\partial_x\psi_L- 
\psi^{\dagger}_R\partial_x\psi_R) \eqno(2)$$
The charge transport through the QPC is affected 
by fluctuations of the charge density
in the 2D reservoirs, which the transitive modes $\psi_{L,R}(x)$ emanate from.
The corresponding fluctuation-induced
electrostatic potential $\Phi(t,{\bf r})$ is governed by the imaginary time 
action   
implied by Eq.(1): 
$$A= \int d\omega\int d{\bf r}\int d{\bf r}^{\prime} \Phi(-\omega,{\bf r})
<{\bf r}|({1\over U_q}+{\sigma_q q^2\over |\omega|}|{\bf r}^{\prime}> 
\Phi(\omega,{\bf r}^{\prime})  \eqno(3)$$
The diffusion-like pole structure of the density correlator (1) implies that 
typical variations of the electrostatic potential occur on length scales
of order $L\sim (k_F\epsilon_0 V/e^2)^{-1/2}$
for an applied voltage bias $V$ (see below). Provided that  
this scale exceeds the linear size of the constriction (which anyway has to be 
under 
the CF mean free path $\sim 1\mu m$),
 one can treat $\Phi(t,{\bf r})$ as an average potential that couples
to a collective phase variable  
$\theta(t)=(2\pi/e) \int^t_{-\infty}dt^{\prime}I(t^{\prime})$
proportional to the total charge passing through the QPC located at ${\bf r
}={\bf 0}$: 
$$\Delta A=\int dt \theta(t) \Phi(t,{\bf 0})  \eqno(4)$$  
By integrating $\Phi(t,{\bf r}\neq {\bf 0})$ out in the standard
fashion, we arrive at the local action for $\theta(t)$:  
$$A_0[\theta] = \int d\omega {|\omega|\over 4g(\omega)} |\theta(\omega)|^2 
\eqno(5)$$
where the effective impedance of the 
electromagnetic environment created by the charge density fluctuations in the 2D 
reservoirs 
is given by the formula  
$$ g^{-1}(\omega)= <{\bf 0}|{U_q\over {|\omega| + \sigma_q q^2 U_q}}|{\bf 0}>  
\eqno(6)$$
The explicit form of the impedance depends on the regime of the 2D electron 
motion
(diffusive or ballistic) and the kind of pairwise potential (Coulomb
or screened). Since we are primarily interested in the strong ballistic 
transmission 
regime, we consider the case of unscreened Coulomb interactions in the absence 
of disorder
which limits our consideration
to frequencies above the bulk impurity scattering rate \cite{6}.
In this regime of interest the momentum-dependent CF conductivity acquires the 
form
$\sigma_q=\nu^2q/(2\pi k_F)$ \cite{8}.

We compute the matrix element (6) in the geometry where the QPC is represented 
by an opening in 
the straight linear 
screen extended along the $y$-axis. The diffusion operator in (6) is to be 
inverted while taking into 
account the "tilted"
boundary condition for the normal component
of the current: $J_x(x=0, y)=\delta(y)$ which vanishes everywhere along the 
screen, besides
at the location of the QPC. As a result we obtain 
$$g(\omega)={\nu}(1-{2\over \ln(E_F/\omega)}+2{\ln\ln(E_F/\omega)\over 
(\ln(E_F/\omega))^2}+ ...) 
\eqno(7)$$
where $E_F\sim e^2k_F/\epsilon_0$ is an upper cutoff set by the CF Fermi energy 
in the bulk.

Neutral quasiparticles described by the operators $\psi_{L,R}$ 
experience the bulk charge fluctuations via the backward scattering term
whose phase includes the transferred charge-counting operator $\theta(t)$ 
\cite{11} 
$$H_{BS}= \lambda{\psi}^{\dagger}_L(0){\psi}_R(0) \exp(i\theta(t)) + h.c. 
\eqno(8)$$

The average current through the QPC is simply related to the voltage drop in the 
constriction 
$I=ge^2/h (V - <\delta V>)$ where $g=g(\omega\rightarrow 0)=\nu$ yields
 the two-terminal d.c. conductance in the limit
 $\lambda\rightarrow 0$, which must be equal to the Hall conductivity in the 
bulk. 

The explicit form of $\delta V$ can be readily found by virtue of the fact that
its integral $\phi(t)=(2\pi e/h)\int^t_{-\infty} dt^{\prime}\delta 
V(t^{\prime})$ 
is a variable canonically
conjugate to $\theta(t)$ ($[\theta, \phi]=2\pi i$):
$$ \delta V = {i\over e}[H_{BS}, \phi]=\lambda (h/e)  Im \psi^{\dagger}_L(0)
\psi_R(0)\exp(i \theta(t)+ I_0t)   \eqno(9)$$
where we separated the fluctuating part of $\theta(t)$ 
from its average value $<\theta(t)>=(2\pi/e)I_0t=(2\pi\nu e/ h)Vt$.

When calculating the current, we first trace out the products of fermion 
operators
$\psi_{L,R}$ whose Fermi-liquid-like matrix elements can be computed with the 
use of Eq.(2)
and absorbed into the definition of $\lambda$.
This way, we arrive at the expression for the backscattering current 
$$I_{BS}=\nu {e^2\over h}\delta V= \lambda \nu e \sin(\theta(t)+{2\pi\over 
e}I_0t)    \eqno(10)$$ 
similar to that in the phenomenological single-mode theory.

This demonstrates that calculation of the conductance    
can be carried out in the framework of 
the effective (zero-dimensional) Caldeira-Leggett-type action 
$$A [\theta]= \int d{\omega}{|\omega|\over 4g(\omega)}|\theta(\omega)|^2 +
\lambda\int dt\cos(\theta(t)+{2\pi\over e}I_0t)
 \eqno(11)$$
Note that, as shown in Ref.\cite{6}, in the opposite, weak tunneling, regime
there exists an alternate description in terms of a conjugate variable 
$\phi(t)$.
The weak-tunneling action of the form (11) contains an
inverted impedance function $1/g(\omega)$, the situation typical for a self-dual 
theory. 

This observation lends a  
support for a tempting proposal of using the single-mode action (11) at 
arbitrary $\nu$ \cite{2}.
For a constant $g=\nu$ the  
exact formulae for the conductance \cite{13} as well as higher correlation 
functions of $I(t)$ \cite{14} 
were obtained by means of the Bethe-ansatz solution. However, it is only this 
case where one can enjoy
the exact solvability, so for a generic $g(\omega)$ one has to resort on 
approximate methods, such
as a perturbative expansion at small $\lambda$. 

We note, in passing, that, although being of an entirely different physical 
origin,
the $\log\omega$ terms in Eq.(7) bear a certain resemblence
to those familiar from the theory of antiferromagnetic spin chains \cite{14.5}
where such terms reflect the presence of a marginally irrelevant operator.
This suggests that it may be possible to recover an effective (1D) sin-Gordon
theory characterized by the bare coupling parameter 
$g=\nu$ from the zero-dimensional (boundary) action (11).  

In the rest of the paper we investigate the deviations from the LL behavior 
corresponding to a constant $g$, which
result from the presence of long-range Coulomb interactions.
The need to include the unscreened Coulomb forces
in order to account for an 
observed departure of the low-temperature conductance of the QPC in the 
$\nu=1/3$ FQHE \cite{15}
from the exact solution of Ref.\cite{13} was pointed out in Refs.\cite{16}. 
The following analysis shows that in the compressible case the effects of 
Coulomb
interactions remain pronounced even in the regime of weak backscattering, unlike 
the situation in FQHE. 

In the second order in $\lambda$ the action (11) yields the backscattering 
current  
$$I_{BS}= \lambda^2{2\pi \nu e\over \Gamma(2\nu)}(\nu eV)^{2\nu -1}(\ln 
{E_F\over \nu eV})^{4\nu}  \eqno(12) $$
which is logarithmically enhanced, as compared to the result for a constant 
$g=\nu$.

The term (12) controls the overall reflection coefficient $R=I_{BS}/I_0$ 
provided that the constriction
opens into the 2D regions adiabatically. This condition  
eliminates backscattering of individual CFs    
from the walls of the constriction and allows the many-body correlation effects 
to dominate
in the conductance reduction.

The perturbative expansion only holds at biases above the crossover $V_{cr}\sim 
\lambda^{1/(1-\nu)}$
 determined by the backscattering amplitude $\lambda$ which one can estimate as 
follows.

Besides some non-universal contribution controlled by the gate potential $V_g$, 
one can also expect an additional, possibly less sensitive to $V_g$, 
contribution
due to the mechanism proposed in \cite{17} where the authors considered
the case of a quantum wire attached to normal, 2D Fermi-liquid-like, reservoirs. 
This extra contribution to $\lambda$ stems from electron scattering off
the static 2D Friedel density oscillations caused by the presence of the 1D 
screen itself. 

Strong correlations present in the CF system make Friedel oscillations decay 
with distance slower
than in a conventional Fermi liquid. 
If, as conjectured by the authors of Ref.\cite{9}, the static $2k_F$-response
exhibits a power-law divergence $\chi(\omega=0,{\bf q})\sim |q-2k_F|^{-\alpha}$,
then the density oscillations decay as 
$$\delta n(x)\sim \int dq_x e^{iq_xx}\chi(\omega =0, q_x, q_y=0)\sim 
\sin(2k_Fx)/x^{1-\alpha} \eqno(13)$$ 
For the exponent falling in the range $1/2<\alpha<1$ 
this gives rise to a singular backscattering amplitude which blows up at 
$k\rightarrow k_F$: 
$$\lambda(k) \sim \int^{\infty}_0rdr\int^{\pi/2}_{-\pi/2}
d\varphi {\cos^2\varphi e^{2ikr}\over r}\delta n(r\cos\varphi)
\sim |k-k_F|^{-\alpha +1/2}   \eqno(14)$$
On the other hand, numerical simulations \cite{10} are rather suggestive of a 
finite 
enhancement of the $2k_F$-response
and a finite $\lambda$ associated with it.

In either case, in contrast to the Fermi liquid situation discussed in 
\cite{17}, 
here we are dealing with the case
$g<1$, so the differential QPC conductance $G=dI/dV$ is likely to vanish as 
$G(V)\sim V^{2/\nu-2}$  for  
biases below $V_{cr}$ (but above the onset of the diffusive
regime \cite{6}), yet arbitrary $V_g$.

Being a potentially by far more informative probe than the conductance, the 
current noise   
 $S(\omega)= \int dt e^{i\omega t} <{\{}I(t), I(0){\}}>$ was recently measured 
in 
 the $\nu=1/3$ FQHE state \cite{18}.
The great deal of interest in this experiment was spurred by the earlier 
theoretical
prediction of a fractional quasiparticle charge, as starkly appearing in the 
zero 
temperature shot noise power \cite{18.5,19,14}
$$ S(0)=2\nu eI_{BS} \eqno(15)$$
In the case of a constriction in a uniform FQHE  
the noise suppression factor can be indeed assigned to a quasiparticle charge  
$Q=\nu e$.
However, this interpretation becomes much less obvious in the case of a QPC 
connecting two
different QH liquids which, according to Ref.\cite{20}, can be formally mapped 
onto 
a uniform QH state with an effective filling factor 
$\nu={2\nu_L\nu_R/(\nu_L+\nu_R)}$.
It is this factor that controls the full transmission current $I_0$    
and also appears in Eq.(15) for the shot noise, but, apparently, it has nothing 
to do with 
the charges of quasiparticles on neither side of the constriction.

For compressible QH states, 
the previous analyses of the current noise produced by CFs in the wide Hall bar 
or annulus geometries
found only minor deviations from the Fermi-liquid results \cite{20.5}.
It is, therefore, of interest to see if measurements of the CF shot noise in 
narrow constrictions
could reveal some new features of these enigmatic quasiparticles.

A systematic perturbative expansion of the shot noise spectrum
in powers of $\lambda$ can be obtained in the framework of the perturbation 
theory
developed in Ref.\cite{19}. In this approach, 
the "glitches" of the phase $\phi(t)\rightarrow \phi(t)\pm 2\pi$
are represented as a plasma
of positive and negative charges confined to the Keldysh contour in the plane of 
complex $t$.

For a generic $g(\omega)$ the charges interact via a 1D potential 
$<\theta(t)\theta(0)>=2\int d\omega
e^{i\omega t} g(\omega)/|\omega|$ that differs from a 1D Coulomb one,
which gives rise to the LL behavior of the averaged products of $\exp(\pm 
i\theta(t))$ operators.
In the case of impedance given by Eq.(7) this results in the  
extra logarithmic factor in 
$D(t)= <e^{i\theta(t)}e^{-i\theta(0)}> \sim t^{-2\nu}(\ln t)^{4\nu}$ 
that had already appeared in Eq.(12).

In the lowest $(\sim \lambda^2)$ order one readily recovers the relation (15).
Thus the interpretation of (15) as an unambiguous manifestation of the 
fractional quasiparticle 
charge has to be taken cautiously, since, apart from the abovementioned
dipolar structure, no physical charge fractionalization is expected for the CFs 
 in compressible QH states. 
On somewhat different grounds, another warning about possible misinterpretations 
of 
the experimentally confirmed relation (15) was recently issued in Ref.\cite{21}.  

We note, in passing, that the abovementioned mechanism of backscattering due to 
the Friedel
oscillations may preclude one from a possibility 
of tuning the QPC into a resonance $(\lambda\rightarrow 0)$ 
by changing $V_g$ and then observing the relation (15) with a doubled prefactor
$(Q=2\nu e)$, as suggested in \cite{18.5}.

In the next $(\sim \lambda^4)$ order
one finds an enhanced negative correction to the ratio $S(0)/(2\nu eI_{BS})$.
In the formalism of Ref.\cite{19},
it stems from "intra-dipole" correlations of the effective 1D charges of 
opposite signs 
separated by time intervals $\Delta t\sim 1/eV$ and can be identified as a 
Coulomb suppression factor $(1-R)$.
Following the arguments of Refs.\cite{18}, in Ref.\cite{22} 
this kind of an additional noise suppression (regardless of its nature)
was identified as an effective quasiparticle charge that varies continuously 
with $R$. 
 Again, this interpretation shows that, being
introduced this way, the notion of quasiparticle charge is more a convention
than a genuine characteristic of the QH state in the bulk.

 Regarding the frequency dependence, in the lowest order 
the current noise is completely flat which corresponds to the "white" Poissonian 
 noise 
without any correlation between different backscattering events.
According to \cite{19},
the first (non-analytical) $\omega$-dependent correction to (15) arises from  
 "inter-dipole" correlations at time scales $\Delta \tau \sim 1/\omega$ (below 
$D^{(n)}=d^nD(\tau)/dt^n$): 
$$\Delta S(\omega)= \lambda^4 (\nu e)^2
[\int tdt e^{i\nu eVt}D(t)]^2 \int d\tau e^{i\omega\tau}
({D^{(2)}\over D} - ({D^{(1)}\over D})^2)\approx {4\pi\over \nu 
e^2}|\omega|({dI_{BS}\over dV})^2  \eqno(16)$$
Focusing on the case of the half filled lowest Landau level $(\nu=1/2)$
we notice that, as opposed to the case of a constant $g=1/2$ \cite{19}, 
the contribution (16) does not vanish, but instead is 
given by the formula
$$\Delta S(\omega)\approx 32\pi|\omega| ({I_{BS}\over eV\ln(2E_F/eV)})^2 
\eqno(17)$$
An experimental observation of this frequency dependence is not 
, however, an easy task. Under the experimental conditions of Refs.\cite{18}
one would have to use frequencies in the $10^2 MHz$ 
range in order to detect $> 1{\%}$ increase in the noise power $S\sim 
10^{-28}A^2/Hz$ 
at typical biases $V\sim 10^2\mu V$
 and currents $I_{BS}\sim 10^2 pA$.
Moreover, when measuring noise in the standard four-terminal geometry one would 
have to correct for
a universal term $\Delta S_0(\omega)=\nu e^2|\omega|/(2\pi)$ \cite{19}.

Besides the noise spectrum, a deviation of $g(\omega)$ from its naive Luttinger 
value 
$\nu=1/2$ would have an effect on such observables as a transresistivity 
$\rho_{12}(T)$ 
measured in the Coulomb drag experiment in the double-layer system \cite{23}.
It was recently suggested \cite{24} that the experimentally observed finite 
transresistivity at
$\nu=1/2$ can be explained as an edge-related effect, based on the calculation 
of $\rho_{12}(T)$
in the system of two locally crossing LLs, each characterized by the parameter 
$g=1/2$.
Although in the LL case the 1D Coulomb drag indeed survives zero temperature 
limit,
the use of the impedance (7) would rather lead to the result that vanishes at 
$T\rightarrow 0$:
$$\rho_{12}(T)\sim (\ln{E_F\over T})^{-4} \eqno(18)$$
Therefore, the observed finite (and even increasing at lower driving
currents) Coulomb drag is more likely to be explained
as a bulk phenomenon caused by strong inter-layer correlations which become  
responsible for a formation of a gapped $\nu=1$ state at smaller distances 
between the layers \cite{25}.

Conceivably, the above analysis permits a natural extension onto incompressible 
FQHE states viewed as   
a system of CFs in a residual effective magnetic field $\Delta B$ \cite{8}. The 
latter provides a
cutoff for the singularities related to the density
response at all momenta $q< \Delta B/k_F$, 
so that an approximate constancy of the impedance function $g(\omega)\approx 
\nu$
and the associated LL behavior follow. We intend to pursue this issue elsewhere.
  
In summary,
we propose a derivation of a single-mode theory of the charge transport through 
a narrow constriction   
 in a compressible QH liquid. Combined with the earlier results \cite{6}
our approach offers a possible explanation of the reported success of an 
empirical fit 
for the data from Ref.\cite{2} by a single-mode Luttinger theory with  
a constant $g\approx \nu$. In addition, we predict 
logarithmic departures from the accustomed Luttinger
behavior, which stem from the long-range  
Coulomb interactions and are expected to 
manifest themselves in the noise spectrum, as well as other observables.

\nopagebreak
  
\end{document}